\documentclass{aa} 
\usepackage{graphicx}
\usepackage{txfonts}
\usepackage{hyperref}

\begin{document} 

\title{Interaction between the ejecta, the accretion disk, and the secondary star in the recurrent nova system U Sco}

\author{Joana Figueira \inst{1,2}
   \and Jordi Jos\'e \inst{1,2}
   \and Rub\'en Cabez\'on \inst{3} 
   \and Domingo Garc\'\i a-Senz \inst{1,2} }

\offprints{J. Jos\'e}

 \institute{Departament de F\'\i sica, EEBE,
            Universitat Polit\`ecnica de Catalunya,
            c/Eduard Maristany 16,
            E-08019 Barcelona,
            Spain\
        \and
            Institut d'Estudis Espacials de Catalunya,
            c/Esteve Terradas 1,
            E-08860 Castelldefels,
            Spain\
        \and
            Center for Scientific Computing (sciCORE), Universit\"at Basel, 
            Klingelbergstrasse 61, 
            CH-4056 Basel, 
            Switzerland
            \\
      \email{jordi.jose@upc.edu}}

\date{\today}

\abstract 
{Recurrent novae are, by definition, novae observed in outburst more than once. They exhibit notably short 
recurrence times between outbursts, ranging from 1 to about 100 yr.  These short recurrence times require very 
high mass-accretion rates, white dwarf masses close to the Chandrasekhar mass limit, and very
high initial white dwarf luminosities.  The likely increase in the white dwarf's mass after
each outburst makes recurrent novae potential type Ia supernova progenitors.}
{Most efforts in the modeling of recurrent novae have centered on the initial phases 
of the explosion and ejection, overlooking the subsequent interaction of the ejecta, first 
with the accretion disk orbiting the white dwarf and ultimately with the secondary star.}
{To address this gap, a series of 3D smoothed-particle hydrodynamics simulations
was conducted. These simulations explored the dynamic interactions between the nova ejecta,
accretion disk, and stellar companion within the framework of the recurrent nova system U Sco.
Notably, the simulations incorporate rotation around the system's center of mass. 
The primary goal of these simulations was to qualitatively examine the impact of various model parameters,
including ejecta mass, velocity, and density, as well as the mass and geometry of the accretion disk.}
{Simulations reveal complete disruption and sweeping of the accretion disk orbiting the white dwarf star 
for models with flared disks and $M_{\rm ejecta}$/$M_{\rm disk} \geq 1$. In contrast, 
V-shaped disks with a (constant) high initial density and $M_{\rm ejecta}$/$M_{\rm disk} < 1$
partially survive the impact with the nova ejecta.
A very minor chemical contamination of the secondary star is anticipated in the U Sco case based 
on the limited impact of nova ejecta particles on the subgiant in all simulations.
Minor mass ejection from the subgiant's outer layers is observed during the late-stage collision with 
ejecta and disk material, with some particles ejected from the binary system and some accreted 
by the white dwarf.}
         {}


\titlerunning{Interaction between the ejecta and the binary system in U Sco}
\authorrunning{J. Figueira et al.} 

\maketitle

\section{Introduction}
Recurrent novae (RNe) are, by definition, novae observed in outburst more than once\footnote{More recently, an alternative criterion, based on the presence of vast nova super-shells,
has also been suggested (see, e.g., the discussion on the putative RN KT Eridani in Shara et al. 2024).}. They exhibit notably short recurrence times between outbursts, 
ranging from 1 yr (M31N 2008-12a; Nishiyama \& Kabashima 2008, Darnley et al. 2016 and references therein) to about a hundred years (e.g., V2487 Oph; Schaefer 2010). 
These short recurrence times require very high mass-accretion rates, white dwarf masses close to the Chandrasekhar limit, and very high initial white dwarf luminosities 
or temperatures (Starrfield, Sparks \& Shaviv 1988, Yaron et al. 2005, Hernanz \& Jos\'e 2008). Moreover, the mass threshold for triggering a thermonuclear runaway in such 
massive white dwarfs is notably lower 
than for lower-mass white dwarfs. This stems from the fact that a more massive white dwarf has a smaller radius and, consequently, a higher surface gravity. Therefore, less mass is required to achieve the critical envelope base pressure, $P_{crit}$ $\sim 10^{20}$ dyn cm$^{-2}$, needed for triggering an outburst (Shara 1981, Fujimoto 1982; see also Wolf et al. 2013 and Kato et 2014,
for more recent work on ignition masses).
Explosions in RNe are often characterized by higher ejection velocities and lower peak luminosities compared to classical novae (Clark et al. 2024).
Theoretical models predict even shorter recurrence times, as brief as 50 days (Wolf et al. 2013, Kato et al. 2014, Hillman et al. 2015), though no observational evidence of such systems has been reported to date. On the other hand, the high mass-accretion rates inferred put constraints on the nature of the stellar companion, requiring an evolved star. 
For instance, the presence of a Roche-lobe-overflowing subgiant star in U Scorpii has been suggested, while mass transfer in RS Oph is expected to result from strong stellar winds emitted by a red giant companion (see Darnley et al. 2012, Darnley 2021, Azzollini et al. 2023).
RNe have garnered interest since observations, as well as simulations, suggest that the mass of the white dwarf component increases after each outburst. Accordingly, these systems 
have been claimed to be likely (type Ia) supernova progenitors (Kahabka et al. 1999, Anupama \& Dewangan 2000, Hachisu \& Kato 2001; see, however, Pagnotta et al. 2015 for a discussion on the 
challenge of determining whether the white dwarf loses or gains mass in U Sco). 

Recurrent novae exhibit a much broader range of orbital periods than classical novae, spanning from a few hours to several hundred days. 
Different subgroups have been proposed, on the basis of different orbital periods but also on the presence of a plateau in the tail of the light curve, different amplitudes
and recurrence times, or different mechanisms driving the high mass-accretion rates (see Schaefer 2010 and references therein). 
A prototype of RNe with intermediate-duration orbital periods is U Sco, which is characterized by
 rapidly declining light curves, with times required to drop two and three magnitudes from maximum brightness of about $t_2$ = $1.2$ days and $t_3$ = $2.6$ days, respectively 
(Schaefer 2010), indicative of the presence of a very massive white dwarf. RNe in this subgroup are further characterized by significantly lower ejected masses (by two orders of magnitude or 
more\footnote{See Caleo \& Shore 2015, and references therein, for estimates of ejected masses in T Pyx, IM Nor, and CI Aql, the known RNe with the shortest orbital
periods. See also Kuin et al. 2020,  for values of the mass ejected in the RN LMC 1968.}) and higher ejection velocities (up to 10000 km s$^{-1}$) compared to classical novae\footnote{Note, however, that recent near-infrared spectroscopy has revealed velocities of the ejecta in U Sco around 22000 km s$^{-1}$, unprecedented in nova outflows (Evans et al. 2023).}. The secondary star in these RN systems is expected to be a subgiant star (Anupama \& Dewangan 2000). 

There is limited knowledge regarding the long-term evolution of these RN systems, 
specifically concerning the interaction of the ejecta with the accretion disk that orbits the white dwarf and, 
ultimately, with the secondary star. Such interactions carry implications for the integrity of the accretion disk (and, therefore, 
for the recurrence period), the amount of material that can
escape the binary system, and the degree of chemical contamination resulting from the impact of a fraction of the nova ejecta 
with the secondary star. This paper, which focuses on the long-term evolution of a specific RN system, U Sco, aims to fill this gap.

The manuscript is structured as follows: The main observational features of U Sco are summarized in Sect. \ref{USco_observables}. The input 
physics and initial conditions adopted are described in Sect. \ref{USco_InputPhysics}.
The impact of rotation, together with a detailed exploration of the parameter space, is presented in 
Sect. \ref{UscoResults}. The most relevant conclusions are summarized in Sect. \ref{UScoConclusions}.

\section{Observational properties of U Sco}
\label{USco_observables}

U Sco (Nova 1863) stands as the third RN identified, among approximately 10 RNe found in the Milky Way\footnote{Several extragalactic RNe have also been discovered in the Large Magellanic Cloud and the Andromeda Galaxy, M31.} (Schaefer 2010).
Located near the northern edge of the Scorpius constellation, at a distance of 19.6$^{+21}_{-5.3}$ kpc from Earth (Schaefer 2018),
U Sco has been seen in outburst in 1863, 1906, 1936, 1979, 1987, 1999, 2010, and most recently on June 6, 2022\footnote{The latest outburst was predicted for the year 2020 $\pm$ 0.7 by Schaefer (2019). Two additional eruptions, in 1945 and 1969, have been suggested by some authors 
(see, e.g., Schaefer 2010).}.
The orbital period of U Sco is approximately 1.23 days (Schaefer 1990, Schaefer \& Ringwald 1995). This relatively long orbital period, coupled with the likely presence of a subgiant companion, implies that accretion in this system may be driven by the expansion of the secondary star toward a red giant configuration, resulting in Roche-lobe overflow. 
This expansion facilitates a sustained accretion at the high rates required from the short recurrence times observed in RNe.

\subsection{The primary star}
The outbursts observed in U Sco are hosted by a massive white dwarf. Although the nature of the primary star (CO- vs. ONe-rich) is still a matter of debate 
(Mason 2011\footnote{See, however, Mason (2013) for 
a reanalysis of the [Ne/O] ratio in U Sco, suggesting only a mild Ne enrichment in the ejecta, possibly consistent with the presence of a CO white dwarf.}), various estimates for its mass have been proposed. Kahabka et al. (1999) suggested M$_{wd}$ $\geqslant$ 1.2 M$_{\odot}$. D\"urbeck et al. (1993) reported M$_{wd}$ $\sim$ 1.16 $\pm$ 0.69 M$_{\odot}$. Hachisu et al. (2000b) assumed M$_{wd}$ $\sim$ 1.37 M$_{\odot}$ to match the quiescent light curve. Finally,  
Thoroughgood et al. (2001) suggested M$_{wd}$ $\sim$ 1.55 $\pm$ 0.24 M$_{\odot}$.
As the total mass of the system likely exceeds the Chandrasekhar limit of 1.4 M$_{\odot}$, U Sco is considered a reliable supernova progenitor 
(Hachisu et al. 2000b, Livio 2000, Thoroughgood et al. 2001; see also Justham \& Podsiadlowski 2008, and Walder et al. 2008, 2010, for a more general discussion on RNe as supernova progenitors).  However, the debate still persists regarding whether the white dwarf mass in U Sco increases or decreases after each outburst (see, e.g., Schaefer 2011, Pagnotta et al. 2015).

\subsection{The secondary star}
The secondary star in U Sco is likely a subgiant (spectral type between F5 and K2; Hanes 1985, Johnston \& Kulkarni 1992, Anupama \& Dewangan 2000), with a mass of 0.88 M$_{\odot}$ and a radius of $\sim$ 2.1 $R_{\odot}$ (Thoroughgood et al. 2001).
Maxwell et al. (2012) reported a helium abundance in the ejecta of U Sco of N(He)/N(H) = 0.117 $\pm$ 0.014, close to solar, suggesting that the subgiant is not helium-rich.  

\subsection{The mass-accretion disk}
The presence of an optically thick, flare-shaped disk has been observationally inferred in U Sco, with a radius of approximately 2.2 R$_{\odot}$ (Schaefer et al. 2011, Mason et al. 2012).
 However, the disk morphology is not well known and some studies suggest the possibility of a raised edge (e.g., Mason et al. 2012).
Overall, the accretion disk is expected to extend up to the inner Lagrangian point of the system, with a radius R$_{\rm disk}$ = 4.21 R$_{\odot}$ during reformation (2.2 R$_{\odot}$ in quiescence; Schaefer 2011). Unfortunately, very little is known about the mass of the disk.

\subsection{The nova ejecta}
Multiwavelength observations of U Sco carried out during its 2010 eruption (Pagnotta et al. 2015), coupled with the empirical relationship between ejected mass and energy radiated during the explosion (Shara et al. 2010), resulted in an estimated ejected mass of $M_{ejec} \sim 2.1 \times 10^{-6}$ M$_\odot$. Other estimates, based on the variation of the orbital period after an outburst, reported different values. For instance, Schaefer (2011)  suggested $M_{ejec} \sim (4.3 \pm 6.7) \times 10^{-6}$ M$_\odot$ for the 1999 explosion, while Schaefer (2013) reported $M_{ejec} \sim 2.5 \times 10^{-5}$ M$_\odot$ for the 2010 outburst. Additional estimates can be found in Hachisu et al. (2000b) and Anupama \& Dewangan (2000).


\begin{figure}[tbh]
  \resizebox{\hsize}{!}{\includegraphics{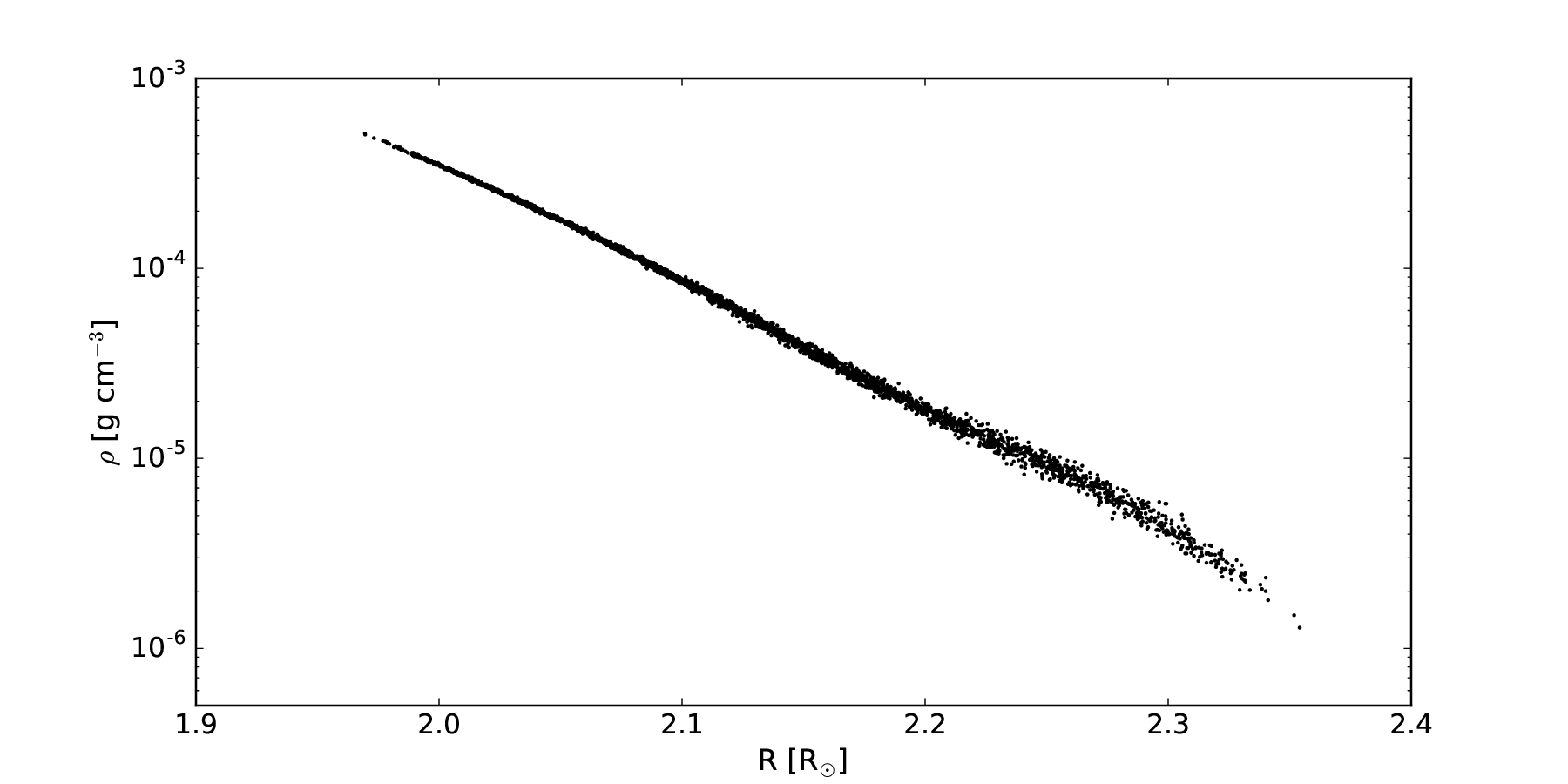}}
            \caption{Density profile for the outer layers of the subgiant secondary star, after relaxation.}
            \label{FigDensity3a}
\end{figure}

\section{Model, input physics, and initial configuration}
\label{USco_InputPhysics}

\subsection{Model}
The methodology employed in this study aligns with the classical nova simulations presented in Figueira et al. (2018; hereafter, Paper I). The accretion, expansion, and ejection stages in U Sco have been modeled with the spherically symmetric, Lagrangian hydrodynamic code 
{\tt SHIVA} (Jos\'e \& Hernanz 1998, Jos\'e 2016). When the inner edge of the ejecta reached escape velocity, at a radius of 0.08 $R_{\odot}$ from the white dwarf center, the structure 
(i.e., mass, radius, density, and temperature) was mapped into a 3D domain, which included the white dwarf star (considered as a point-like mass), 
the mass-accretion disk, and the secondary star. The subsequent evolution of the system was then followed with the 3D, smoothed-particle hydrodynamics (SPH) code {\tt GADGET-2} (Springel, Yoshida \& White 2001, Springel \& Hernquist 2002, Springel 2005). 
 To determine the physical conditions that govern the stability of the accretion disk in a RN system like U Sco, we examined the specific impact of various model parameters, including 
ejecta mass, velocity, and density, as well as the mass and geometry of the accretion disk. To properly characterize the role played by each of these parameters, we carried out a series of 
simulations that differ only in the value adopted for one of these parameters. Therefore, while Model C was evolved with a nova ejecta that closely corresponds to that obtained 
in our 1D simulations with {\tt SHIVA}, we varied the conditions in the other models by assuming, for example, a less massive ejecta (Model B), a higher or lower velocity ejecta (Models A and D), 
a more massive accretion disk (Model E), and different geometries (flared vs. V-shaped) and densities (profile extracted from {\tt SHIVA} vs. constant density) of the disk. 

An orbital period of $P_{\rm orb} = 29.53$ hr (1.23 days), 
characteristic of U Sco, has been adopted in all the simulations reported in this paper\footnote{This period is $\sim$ 250 times longer than the time required for the ejecta to reach and hit the secondary star, at a speed of 10000 km s$^{-1}$.}. 
This orbital period is about 3.3 times larger than the one adopted in Paper I for classical novae, $P_{\rm orb} = 8.9$ hr. 
In addition,  orbital rotation around the center of mass of this wider binary system has been included in this paper. 

\subsubsection{The primary star}
The white dwarf is modeled as a point-like mass of 1.38 $M_{\odot}$, 
which is sufficient to account for the gravitational pull exerted by the star on the overall system. 

\subsubsection{The secondary star}
The secondary star is represented by a subgiant companion of 0.88 $M_\odot$ and solar metallicity. The structure of the secondary was computed in spherical symmetry using the {\tt FRANEC} code (M. Salaris, private comm.). Upon entering the subgiant branch, the secondary reaches a radius of about 2.2 $R_\odot$ and a luminosity of 
Log $L(L_\odot) \sim 0.343$ (Log $T_{eff} = 3.69$). The specific values align with those inferred for the companion star in U Sco as reported by 
Thoroughgood et al. (2001). 
The envelope was subsequently mapped onto a 3D particle distribution,  although only the outermost layers (i.e., the outer $0.0054$ 
M$_\odot$) were finally taken into account to avoid an excessive computational load.
The SPH mass particles of the secondary star move rigidly around the center of mass of the system with the same angular velocity.
Given that particles from the ejecta are not expected to penetrate deeply into the secondary star, 
its innermost layers, those not directly resolved by the simulation, have been replaced by a point-like mass located at the center of the star.
To generate the initial 3D particle distribution (density) of the outer layers, a glass technique has been implemented (White 1996; see also Paper I). The resulting density profile, after relaxation, is shown in Fig. \ref{FigDensity3a}. The radius of the star, after relaxation, reached a value of 2.37 $R_\odot$.

\subsubsection{The mass-accretion disk}
\begin{figure*}[bth]
\centering
\includegraphics[height=5truecm]{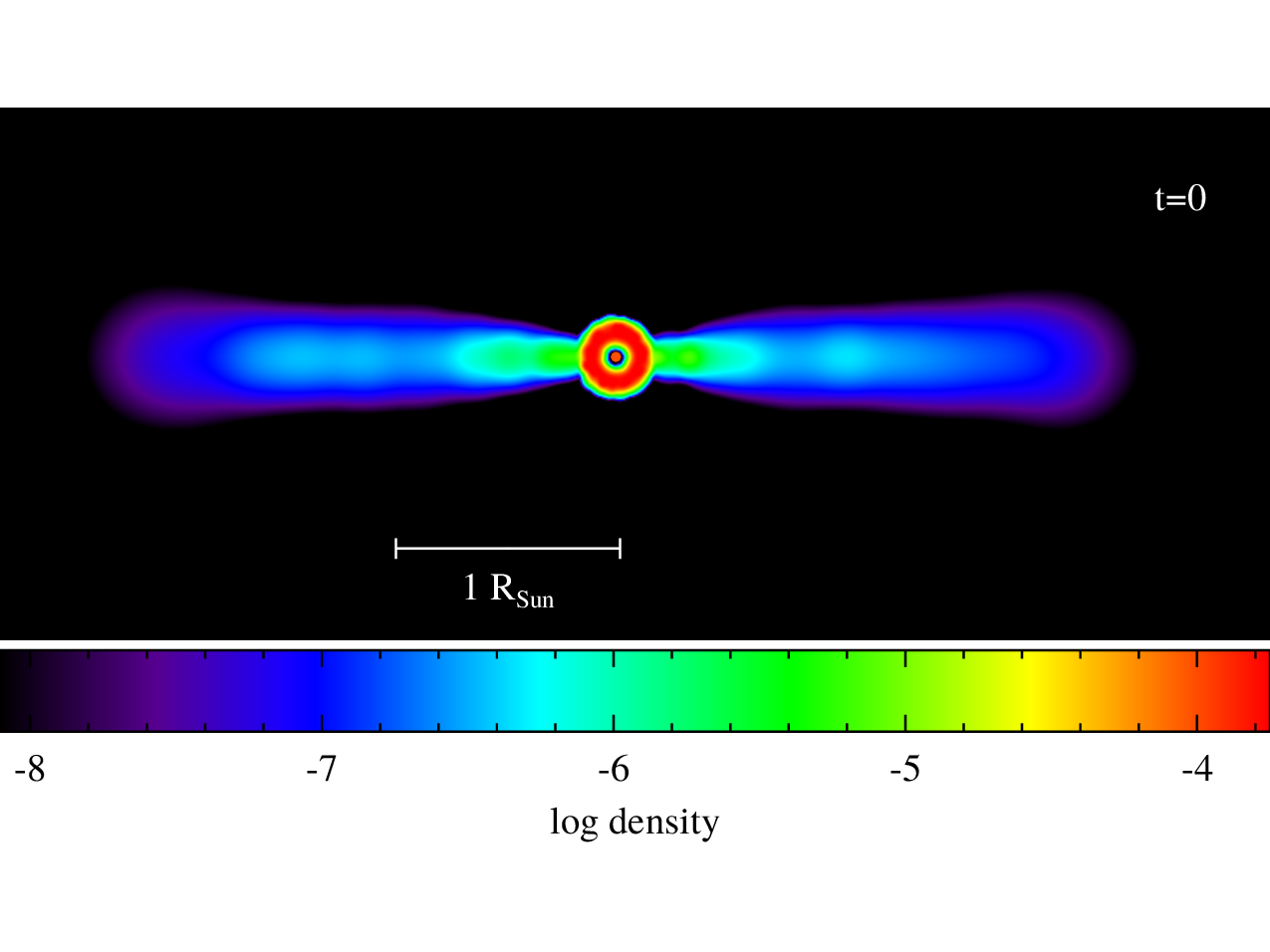} \label{Vshaped}
\includegraphics[height=4.7truecm]{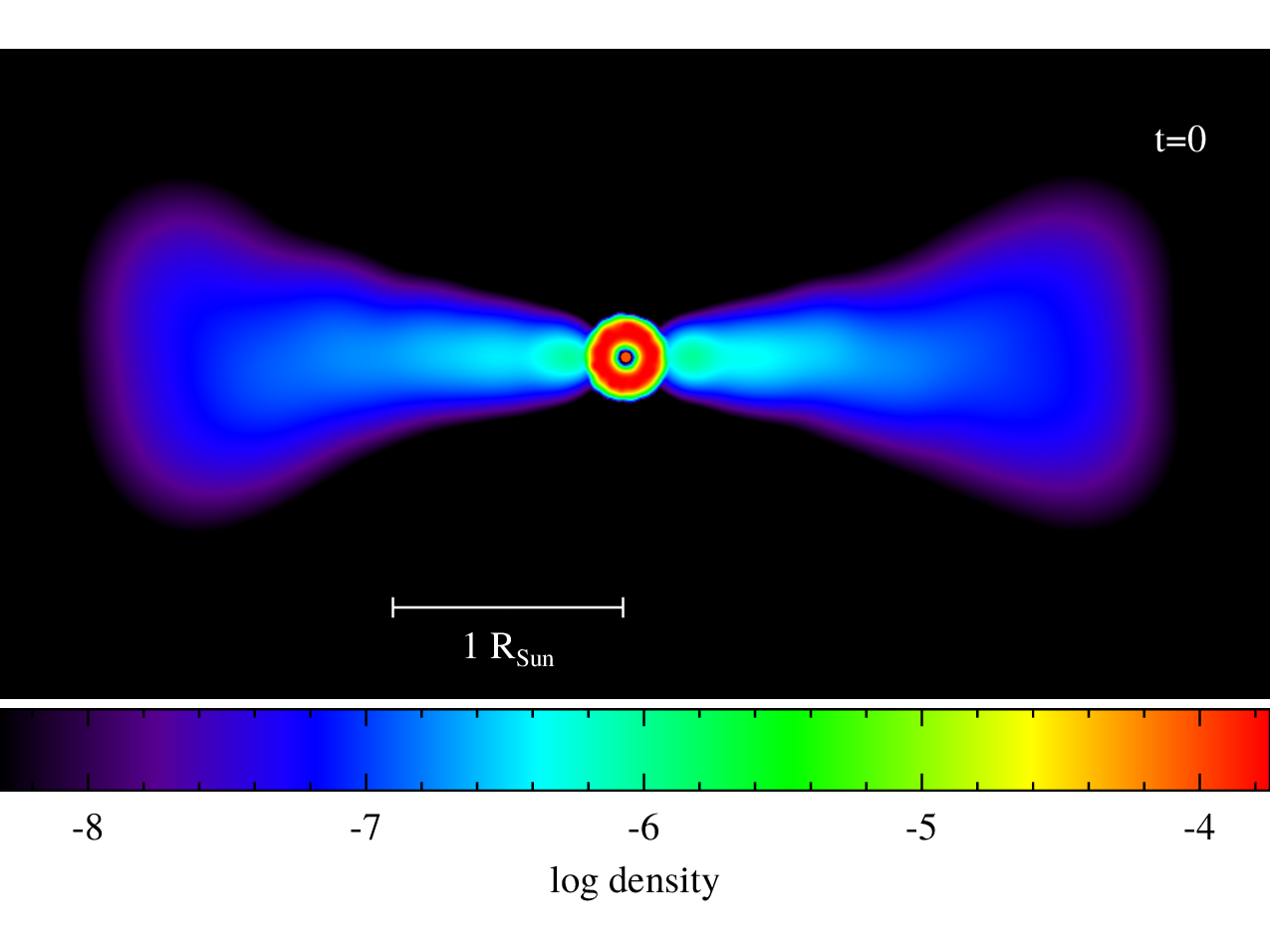} \label{Flared}
\caption{Different geometries for the mass-accretion disk: 
a V-shaped disk (left panel) and a flared disk (right panel).}
\label{Flared_VShaped}
\end{figure*}

\begin{figure}[bht]
  \resizebox{\hsize}{!}{\includegraphics{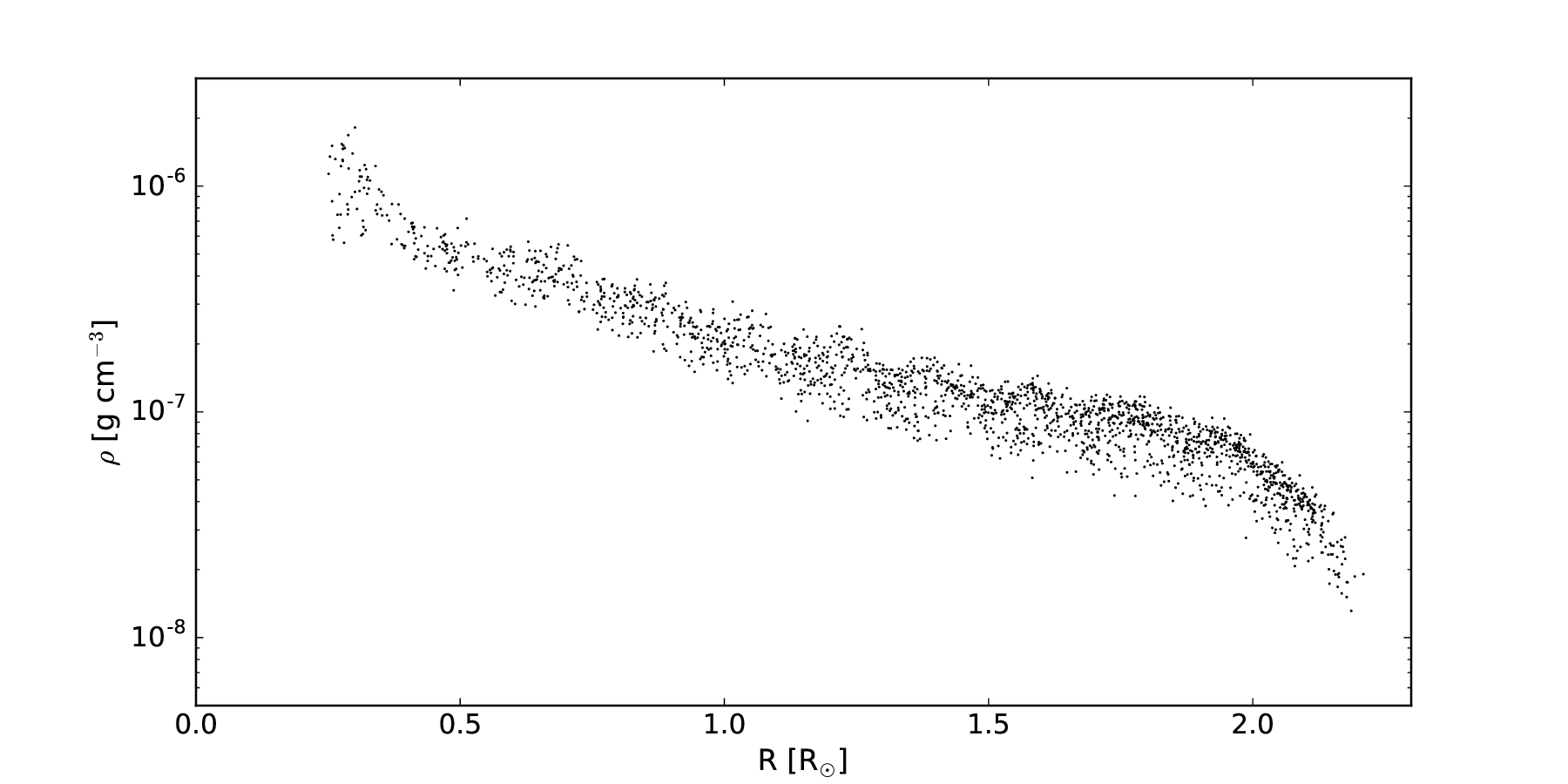}}
            \caption{Initial density profile for the mass-accretion disk.}
            \label{FigDensity3b}
\end{figure}

A flared mass-accretion disk (see Fig. \ref{Flared_VShaped}) that orbits the white dwarf in Keplerian rotation has been adopted, following the prescription outlined by Hachisu et al. (2000a). 
Estimates of the disk density have been obtained from the Shakura \& Sunyaev (1973) model.
In our fiducial model (hereafter, Model A), a solar-composition disk was adopted, with a mass of $10^{-6}$ $M_\odot$ (similar to the mass of the ejecta) 
and a geometry given by
\begin{equation}\label{EqDISK}
H = \beta R_{disk} \left(\frac{d}{R_{disk}}\right)^{\nu},
\end{equation}
where $H$ represents the height of the disk from the equatorial plane, $d$ is the distance from the white dwarf’s center, $\nu$ is an exponent related to the surface shape (set as $\nu = 2$, based on Hachisu et al. 2000b\footnote{The exact value of $\nu$ has no significant effect on the results (Hachisu et al. 2000b).}),  
and $\beta$ is a parameter indicating the disk’s thickness (a value of $\beta = 0.3$ has been adopted, following Hachisu et al. (2000b)).
The initial density profile for the mass-accretion disk is depicted in Fig. \ref{FigDensity3b}. 
For comparison (see Sect. \ref{GeoDisk}), a V-shaped disk (see Paper I for details) has also been adopted in some of the models computed.

\subsubsection{The nova ejecta}
\begin{figure*}[bth]
\centering
\includegraphics[height=7truecm]{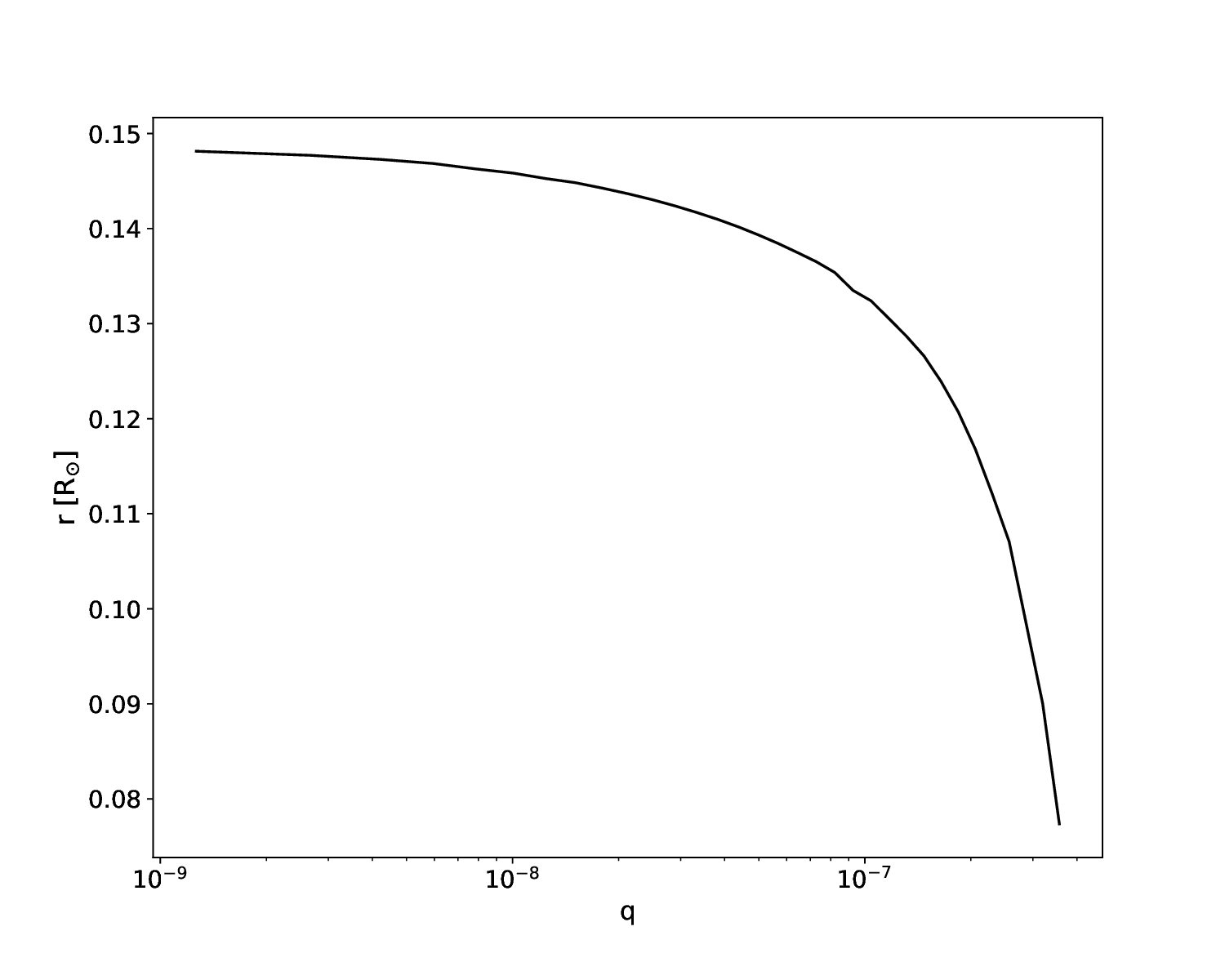} 
\includegraphics[height=7truecm]{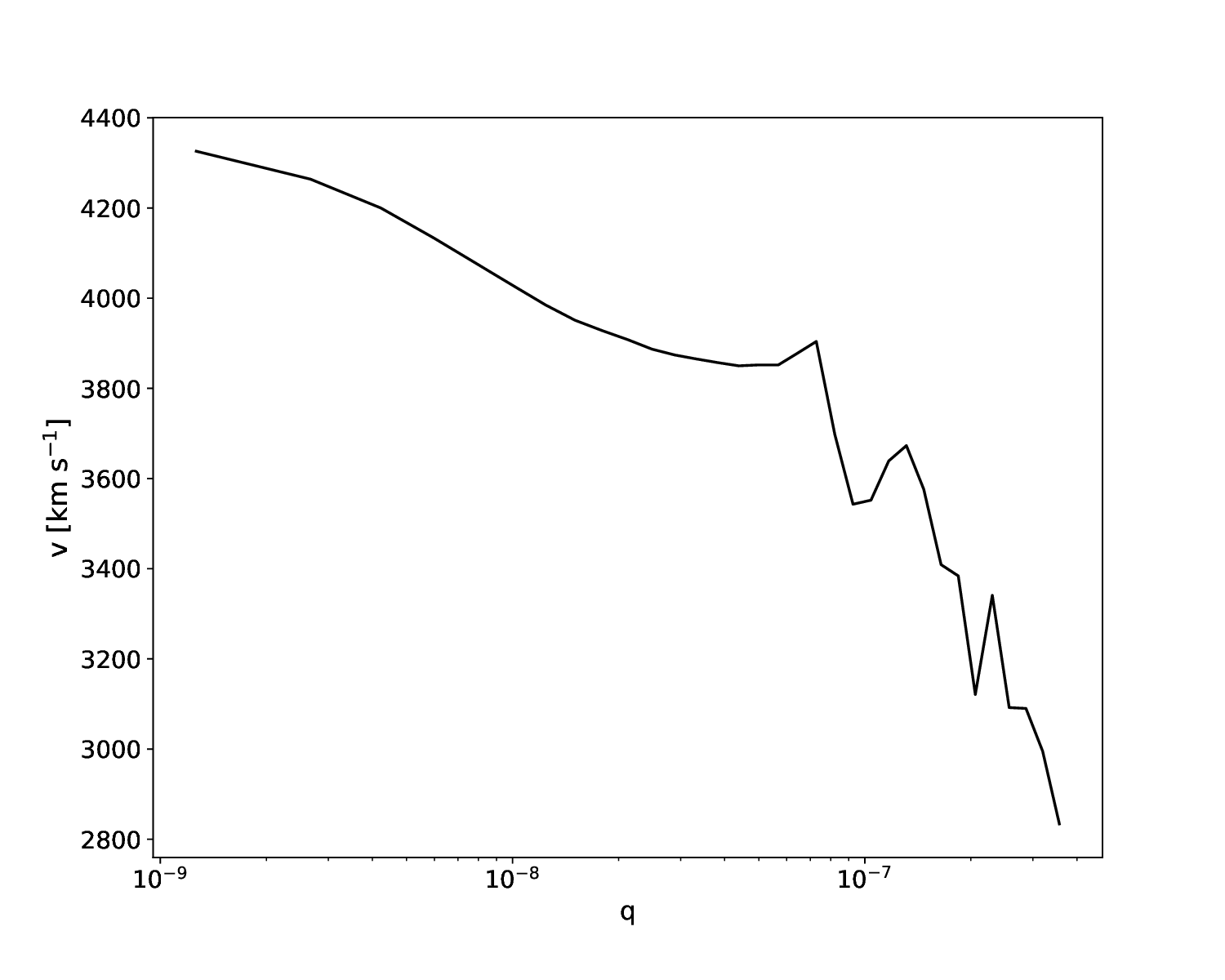} 
\includegraphics[height=7truecm]{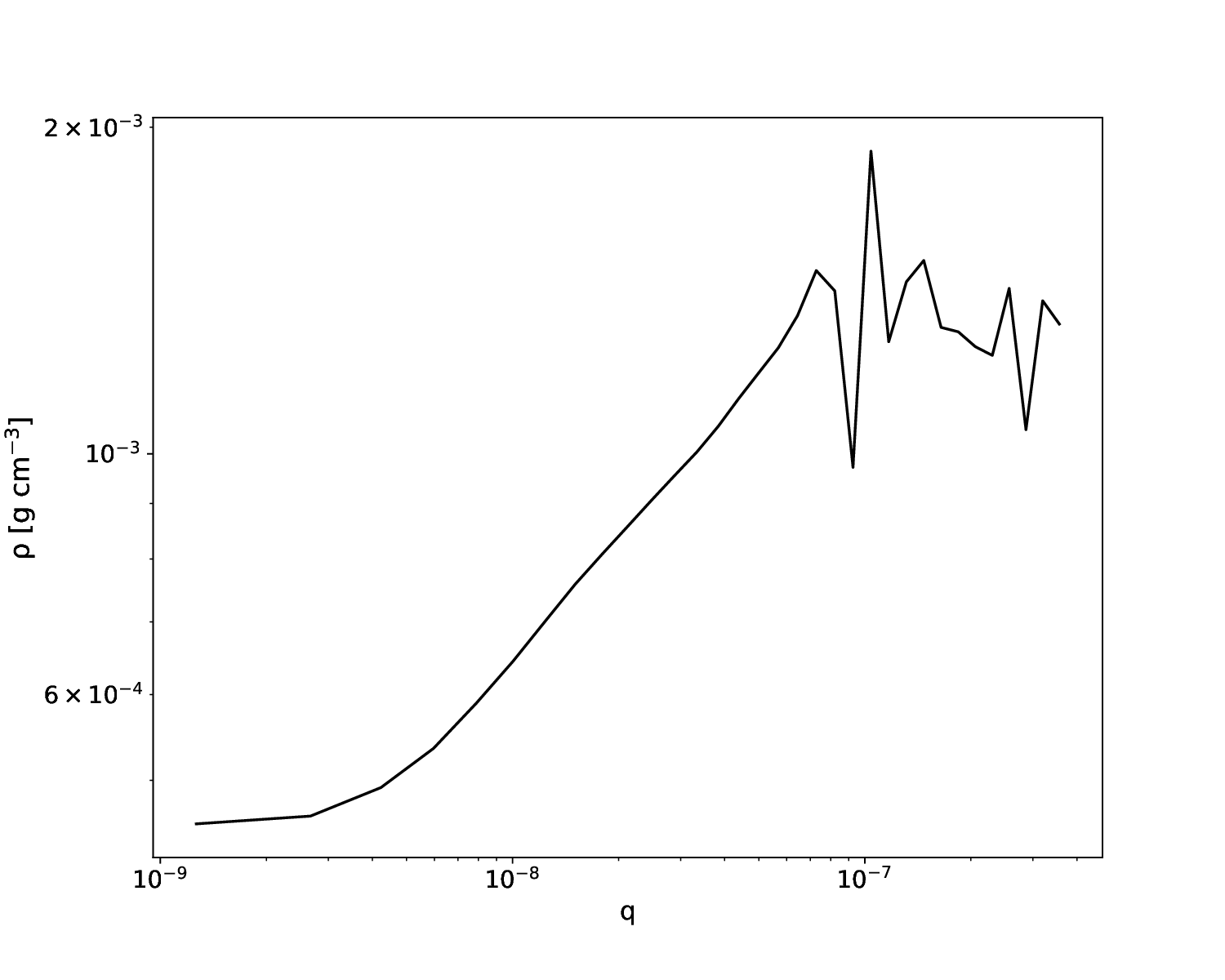} 
\includegraphics[height=7truecm]{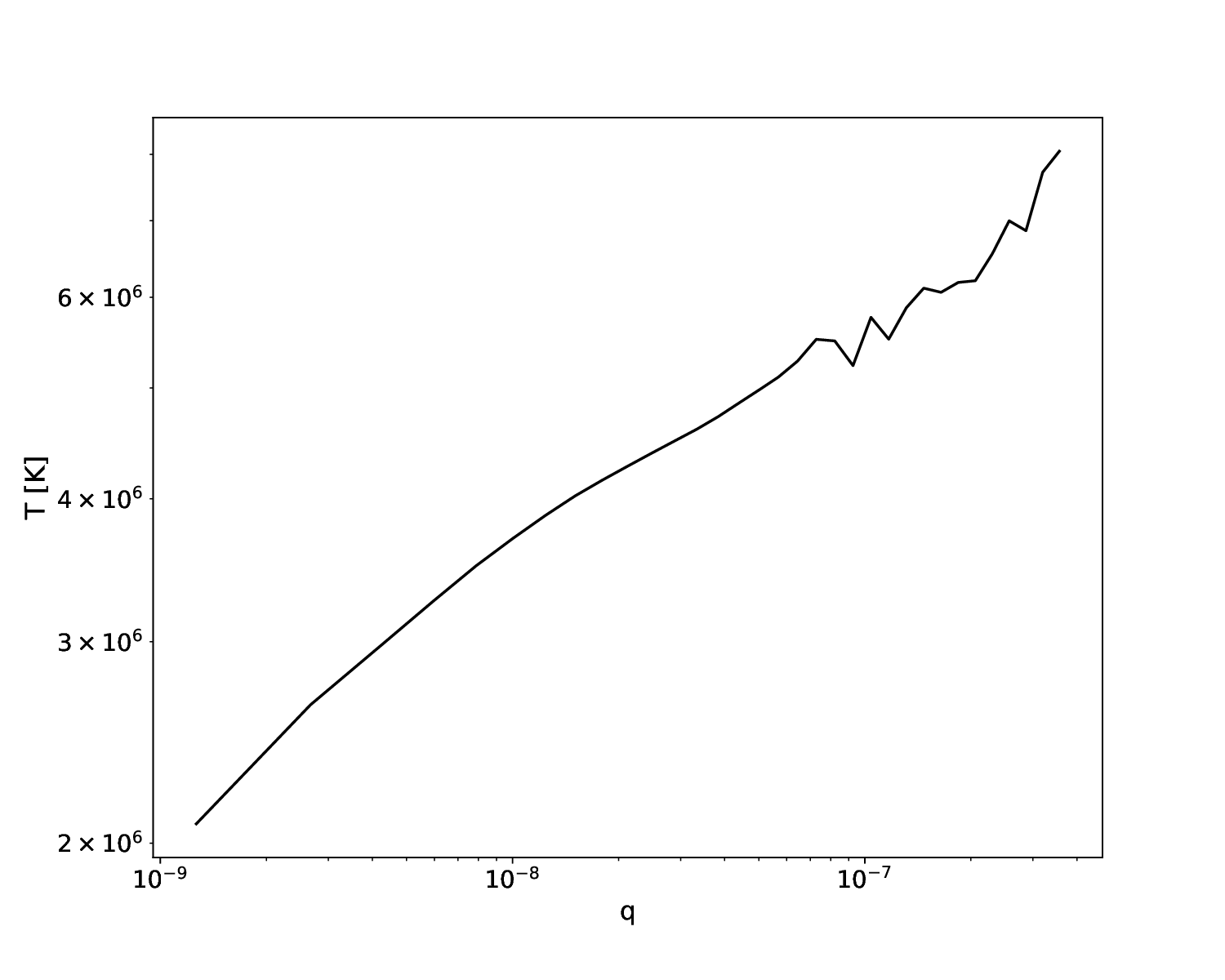} 
\caption{Radius (upper left panel), velocity (upper right), density (lower left), and temperature profiles (lower right) of the ejecta as computed with the 1D code {\tt SHIVA}. 
         The mass coordinate is evaluated as q = 1 - m$_{int}$/M$_{wd}$.}
\label{SHIVAini1}
\end{figure*}

The expanding nova ejecta, initially located between 0.08 $R_{\odot}$ (inner edge) and 0.15 $R_{\odot}$ (outer edge) from the underlying white dwarf, has a mass of $2 \times 10^{-6}$ $M_\odot$, a mean metallicity of $Z = 0.011$ (i.e., solar), and a density and velocity profiles (Fig. \ref{SHIVAini1}) matching the values obtained with the 1D code {\tt SHIVA}. 

\subsection{Resolution}

To ensure computational accuracy and facilitate precise interpolation, all SPH particles employed in the simulations have the same mass, $5 \times 10^{-10}$ $M_\odot$.  In the models reported in this paper, a total of 9.77 million SPH particles have been employed. 
 The majority of these particles are allocated to 
the outer layers of the secondary star due to its comparatively higher mass, while the much lighter accretion disk and  nova  ejecta  only 
comprise 2000 and 3900 particles, respectively, which represents the limiting factor of the simulations.

\subsection{Relaxation of the initial model}
A standard relaxation procedure has been applied to the SPH particles to ensure that the 
 3D structure is in hydrostatic equilibrium before the simulations with {\tt GADGET-2} begin.  
Similar to the models described in Paper I, the initial relaxation is performed on the stellar secondary in isolation, spanning a duration equivalent to 8 orbital periods. In the new 
simulations reported here, a second relaxation stage has been implemented, following the procedure described in Rosswog et al. (2004) and Lajoie \& Sills (2010): 
both the primary and the secondary star are placed in the co-rotating reference frame of the binary system, for about two additional orbital periods. Once the system reaches a stable 
configuration, we move the system back to the reference laboratory frame of the center of mass and add the ejecta and the disk encircling the primary.
This dual-stage relaxation aims to enhance the stability of the binary system, mitigating the initial oscillations reported in Paper I.

The sound-crossing time throughout the disk is $\sim$ 5 hours, while the time required for the ejecta to reach and impact the disk is just $\sim$ 16 seconds. 
Consequently, no explicit relaxation for the disk, 
also generated using the glass technique, is deemed necessary.

\section{Results}
\label{UscoResults}

\subsection{Evolution of Model A}
\begin{figure*}[bth]
            \centering
            \includegraphics[width=\textwidth]{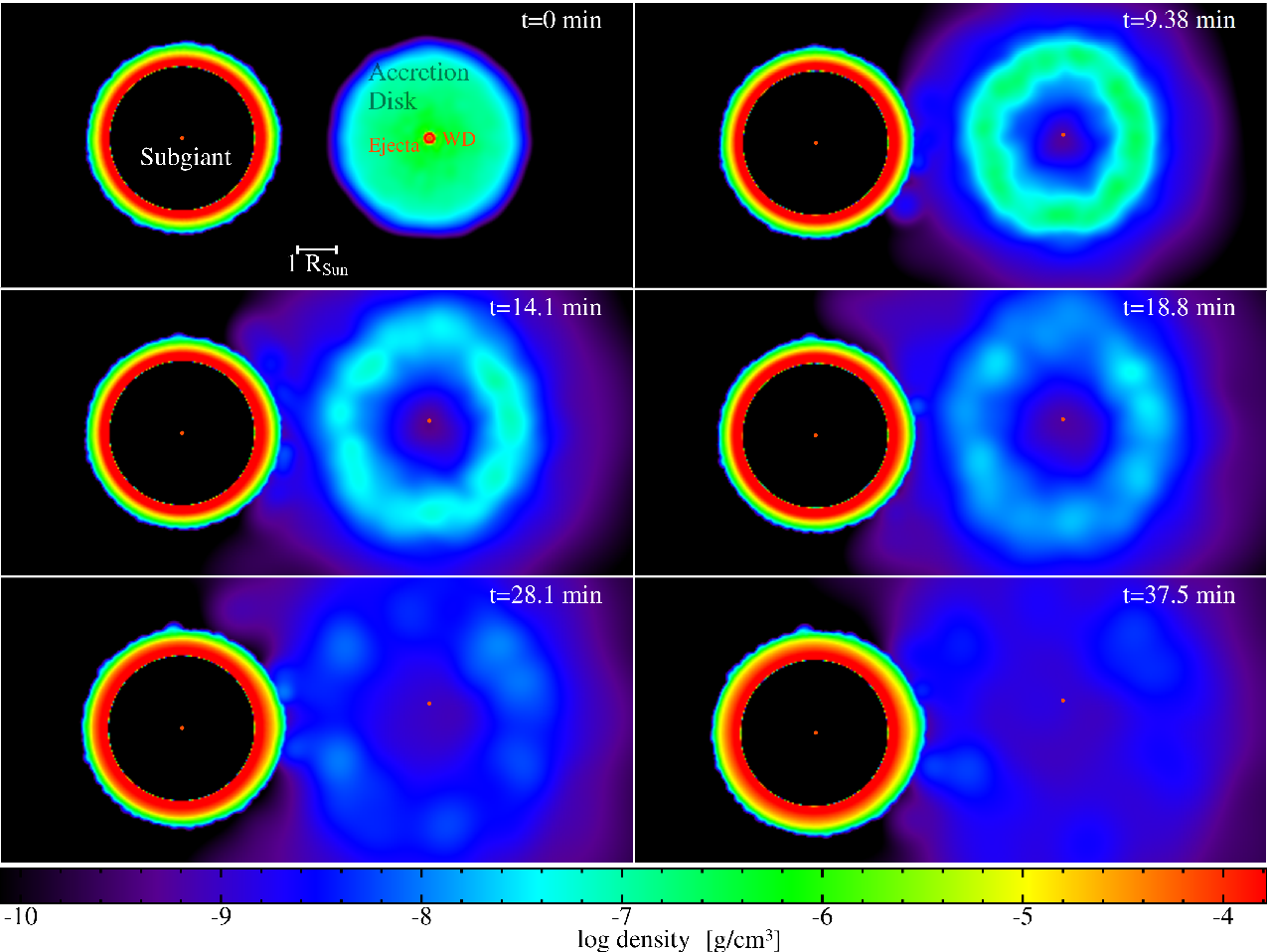}
            \caption{Cross-sectional slices showing the density distribution within  the binary orbital plane (XY) in Model A at various stages of the interaction between the nova ejecta and the mass-accretion disk, followed by a collision with the subgiant stellar companion. 
A movie showcasing the time evolution of this model is available online.
Both the snapshots and the movie were generated using the visualization software {\tt SPLASH} (Price 2007).}
            \label{evolutionAA}
        \end{figure*}

Model A (see Table \ref{Series2a}) describes the interaction between the material ejected during a RN outburst and the surrounding environment. Specifically, 
 the model assumes $M_{\rm ejecta} =
2.1 \times 10^{-6}$ $M_\odot$ ejected from a 1.38 $M_\odot$ white dwarf during the outburst, with a maximum velocity of $V_{\rm ejecta}^{\rm max} = 10000$ km s$^{-1}$ 
and a density profile calculated with the {\tt SHIVA} code. 
The ejecta hits a $10^{-6}$ $M_\odot$ flared, accretion disk (Fig. \ref{Flared_VShaped}), and subsequently, the outer layers of the 0.88 $M_\odot$ subgiant companion (see Tables \ref{Series2a} and \ref{Series2b}). 
Snapshots capturing the dynamic evolution of Model A are shown in Fig. \ref{evolutionAA}. The time evolution of the density of the system is depicted in 
a movie available online\footnote{Movies showing the evolution of several models are also available at \url{http://www.fen.upc.edu/users/jjose/Downloads.html}}.

\begin{figure}[bth]
  \resizebox{\hsize}{!}{\includegraphics{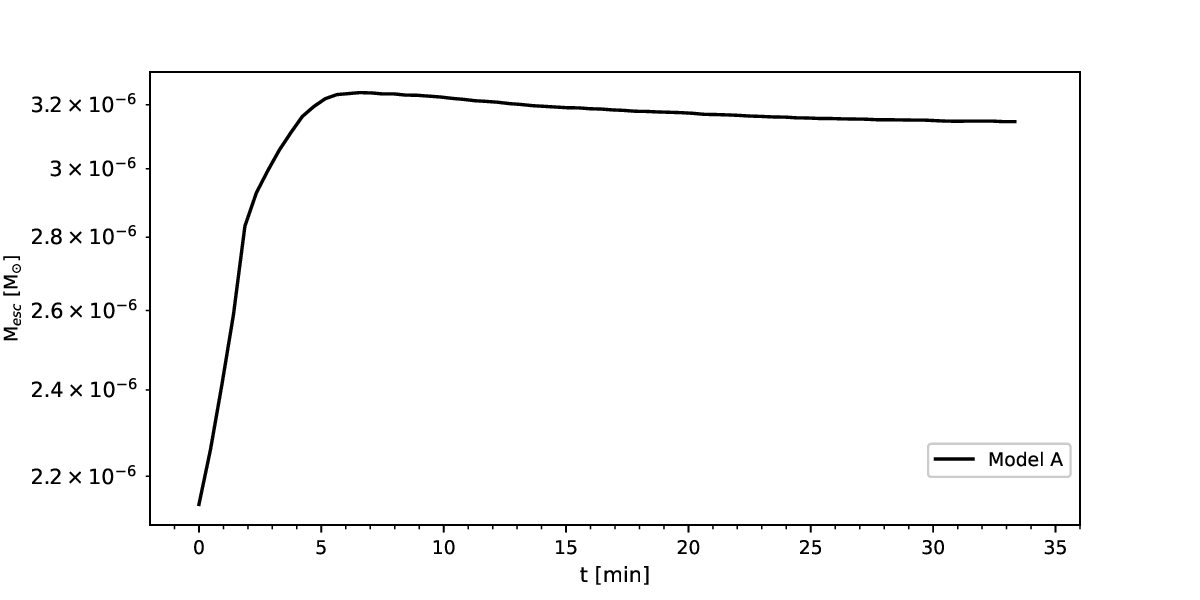}}
            \caption{Time evolution of the mass leaving the binary system in Model A.}
            \label{FigMlossA}
\end{figure}

The high ejection velocities drive the rapid collision between a fraction of the ejected plasma and the mass-accretion disk, occurring $\sim$ 16 s after the simulation begins (Fig. \ref{evolutionAA}, upper panels). 
The heat released in this energetic collision raises the disk's temperature to an average of $\sim 2.6 \times 10^7$ K for a few minutes\footnote{At the end of simulations
reported for Model A ($t \sim 40$ min), the particles from the disk have already cooled down to an average temperature of $\sim 1.5 \times 10^6$ K.}, while a few hundred particles, 
constituting about 4\% of the total disk plus ejecta particles, exceed $10^8$ K. The low plasma density, $\sim 3 \times 10^{-5}$ g cm$^{-3}$, suggests that nuclear processing would not substantially  alter its composition.
Despite U Sco's wide orbit, about 20\% of the ejecta collides with the mass-accretion disk, $m'_{\rm ejecta} = 0.2 M_{\rm ejecta}$, due to the large cross-sectional area of the adopted flared geometry (see Fig. \ref{Flared_VShaped}). The mean kinetic energy of this ejecta, $K = \frac{1}{2} m'_{\rm ejecta} V_{\rm ejecta}^2 \sim 4 \times 10^{44}$ ergs, is larger than the gravitational binding energy of the disk, $U \sim G M_{\rm WD} M_{\rm disk} / r_{\rm mean} \sim 3 \times 10^{42}$ ergs ($G$ is the gravitational constant, $M_{\rm WD}$ is the white dwarf mass,  and $M_{\rm disk}$ and $r_{\rm mean}$ are the disk mass and the mean distance between the white dwarf and the disk\footnote{A value of 2.12 $R_\odot$ (i.e., the outer radius of the disk) has been adopted for $r_{\rm mean}$. This choice clearly overestimates the binding energy of the disk, which is still smaller than the mean kinetic energy of the ejecta.}), resulting in the complete disruption of the disk in Model A. 

At $t \sim 7$ min (middle panels), a mixture of ejecta and disk material impacts the subgiant companion. This collision induces a modest temperature increase in the outer layers of the star, reaching a mean value of $\sim 3.5 \times 10^6$ K. The energy released prompts a moderate expansion of the outer layers of the subgiant (lower panels). Despite those effects\footnote{While part of the
energy deposited in the outer layers of the secondary by the impact with the ejecta plus disk particles will be radiated away, a fraction will be transported inward. An accurate analysis
of the observability of this transient event would require the use of a stellar evolution code to properly address the time-evolution of the subgiant star, which is beyond the scope of the
present work.}, the moderate temperatures achieved suggest that nuclear processing can be disregarded.

About $\sim 3.1 \times 10^{-6}$ $M_\odot$ (roughly 97\% of the combined disk and nova ejecta) leave the binary system in Model A. 
In contrast, only $\sim 9.8 \times 10^{-8}$ $M_\odot$ (about 3\%) becomes gravitationally bound to the white dwarf (see Table \ref{Series2b}). 
The collision results in a minimal incorporation of ejecta/disk particles into the outer layers of the subgiant companion: only 4 out of  the 29 SPH particles that impacting secondary (up to 37 minutes into the simulation) become part of the star, representing a total mass of $2.2 \times 10^{-9} M_\odot$, or $\sim$ 0.07\% of the overall disk and nova ejecta (a small amount, as expected from the large orbital period adopted and the large separation between the two stars, as well as the large velocities and small masses that characterize the ejecta in this model\footnote{Similar patterns have been found in all the simulations reported in this paper.}). However, it is worth noting that these results are qualitative and a detailed account of the contamination of the secondary star would require a much larger number of SPH particles for the disk plus ejecta mixture.

Figure \ref{FigMlossA} shows the time evolution of the mass leaving the binary system in Model A. 
The early and pronounced increase in mass loss (up to 7 min) is attributed to the interaction between the nova ejecta and the disk, which leads to complete sweeping and mixing. 
The subsequent evolution reveals limited mass loss from the outer layers of the subgiant companion, with a minor decline in M$_{\rm esc}$ due to particles achieving marginal escape velocity and subsequent deceleration by interactions with material orbiting the binary system.

It is worth noting that rotation plays a minor role for a binary system with U Sco's wide orbit. The effect of rotation has been assessed with a test simulation in which 
rotation was turned off, revealing small differences ($<$ 1\%) in the fraction of material that leaves the system or remains gravitationally bound to the white dwarf over
the total ejecta plus disk mass.   
However, more significant contrast was found in the fraction of mass lost by the subgiant that ultimately becomes gravitationally bound to (accreted by) the white dwarf, $\Delta M_{\rm SS, WD}$, 
about 20\% smaller when rotation was considered. 

   \begin{table*}[thb]
   \caption{U Sco models computed.}
   \label{Series2a}
   \centering                         
   \begin{tabular}{c c c c c c }   
   \hline\hline\
   Model& $M_{\rm ejecta}$& $V_{\rm ejecta}^{\rm max}$& $\rho_{\rm ejecta}$& $M_{\rm disk}$& Shape Disk\\ 
        &      ($M_\odot$)&  (km s$^{-1}$)            &                  &  ($M_\odot$)  &           \\ 
   \hline
   A & $2.1 \times 10^{-6}$& 10000& {\tt SHIVA}  & $10^{-6}$           & Flared \\ 
   B & $1.1 \times 10^{-6}$& 10000& {\tt SHIVA}  & $10^{-6}$           & Flared \\
   C & $2.1 \times 10^{-6}$& 5000 & {\tt SHIVA}  & $10^{-6}$           & Flared \\ 
   D & $2.1 \times 10^{-6}$& 2000 & {\tt SHIVA}  & $10^{-6}$           & Flared \\ 
   E & $2.1 \times 10^{-6}$& 10000& {\tt SHIVA}  & $4.4 \times 10^{-6}$& Flared \\ 
   F & $2.1 \times 10^{-6}$& 10000& Const. & $10^{-6}$           & V-Shaped \\
   G & $2.1 \times 10^{-6}$& 10000& Const. & $10^{-6}$           & Flared \\ 
   H & $2.1 \times 10^{-6}$& 10000& {\tt SHIVA}  & $10^{-6}$           & V-Shaped \\ 
    \hline
    \end{tabular}
    \end{table*}

\subsection{Exploration of the parameter space}
The inclusion of rotation, varied disk shapes, and  distinct physical properties of the ejecta in RNe --- such as lower ejected masses with higher velocities 
compared to classical novae and different densities --- calls for a  reanalysis of the interaction between the nova ejecta, the accretion disk and the companion star within a wider binary system like U Sco ($P_{\rm orb} = 29.53$ hr). 
Specifically, we focused on the crucial parameters influencing the long-term evolution of the system, including ejecta mass, velocity, and density, as well as the mass and geometry of the accretion disk (see Tables \ref{Series2a} and \ref{Series2b}).

It is worth noting that models with flared disks and ratios $M_{\rm ejecta}$/$M_{\rm disk} \geq 1$ (i.e., Models A, B, C, D, G, and H) 
 result in complete disruption of the disk during the nova blast.
In contrast, in Model E, characterized by a lower $M_{\rm ejecta}$/$M_{\rm disk} = 0.48$ (indicating a massive disk), 
the disk partially survives the collision (note also that different density ratio between the outer and the inner disk regions in Model D, $\sim 12$, and in Model E, $\sim 3$).
Another noteworthy case is Model F, where a V-shaped disk featuring a constant, high initial density (in sharp contrast to Model H, characterized also by a V-shaped disk, but with a density profile similar to the one depicted in Fig. \ref{FigDensity3b}), endures the collision with the ejecta.

\subsection{Effect of the mass of the nova ejecta}
To assess the impact of the ejecta mass ($M_{\rm ejecta}$) on the simulation outcomes, two different values have been considered, as detailed in Table \ref{Series2b}. A comparison between our fiducial Model A (with $M_{\rm ejecta}$ = $2.1 \times 10^{-6}$ $M_\odot$) and Model B ($1.1 \times 10^{-6}$ $M_\odot$) reveals that a more massive ejecta, characterized by a greater momentum,
 results in a greater mass leaving the binary system. In turn, a lower mass remaining gravitationally bound to the white dwarf is found.  
This marks a significant departure from the findings in Paper I.
In this work, the adoption of larger velocities for the ejecta, a characteristic feature of RNe, with $V^{\rm max}{\rm ejecta}$ = 10000 km s$^{-1}$, results in a substantial 
fraction of the disk and ejecta mass exceeding the escape velocity from the system following the collision, reducing the amount of matter that gets bound to the white dwarf.
The subsequent interaction of the ejecta plus disk with the subgiant star is also affected. 
Specifically, Table \ref{Series2b} shows a decrease in the mass ejected from the system by the secondary ($\Delta M_{\rm SS, lost}$) 
and a reduction in the fraction of mass ultimately accreted by the white dwarf ($\Delta M_{\rm SS, WD}$) as the ejecta mass increases.

\begin{table*}[t]
\caption{Results of the models computed.}
    \label{Series2b}
    \centering   
    \begin{tabular}{c c c c c c c c c}
    \hline\hline\
    Model& $\Delta M_{\rm WD}$($M_{\odot}$) & $\Delta M_{\rm esc}$($M_\odot$)   &
    $\frac{\Delta M_{\rm WD}}{(M_{\rm ejecta} + M_{\rm disk})}$  & $\frac{\Delta M_{\rm esc}}{(M_{\rm ejecta} + M_{\rm disk})}$ & 
    $\Delta M_{\rm SS, lost}$($M_{\odot}$)  & $\Delta M_{\rm SS, WD}$($M_{\odot}$)  & Disk disruption
\\
    \hline
    A    & $9.84 \times 10^{-8}$ & $ 3.14\times 10^{-6}$ & 3.04\%  & 96.96\% &$5.5 \times 10^{-10}$& $5.92 \times 10^{-7}$& Yes  \\
    B    & $1.61 \times 10^{-7}$ & $ 2.14\times 10^{-6}$ & 7.01\%  & 92.99\% &$1.1 \times 10^{-9} $& $2.37 \times 10^{-6}$& Yes \\
    C    & $1.92 \times 10^{-7}$ & $ 3.05\times 10^{-6}$ & 5.94\%  & 94.06\% &$2.75 \times 10^{-9}$& $1.93 \times 10^{-6}$& Yes \\
    D    & $2.96 \times 10^{-7}$ & $ 2.95\times 10^{-6}$ & 9.14\%  & 90.86\% &$1.1 \times 10^{-9} $& $2.01 \times 10^{-6}$& Yes \\
    E    & $1.95 \times 10^{-6}$ & $ 4.58\times 10^{-6}$ & 29.87\% & 70.13\% &$1.58 \times 10^{-8}$& $2.22 \times 10^{-6}$& Partially  \\
    F    & $8.69 \times 10^{-7}$ & $ 2.35\times 10^{-6}$ & 26.97\% & 73.03\% &$3.85 \times 10^{-8}$& $9.41 \times 10^{-7}$& Partially \\
    G    & $3.46 \times 10^{-7}$ & $ 2.89\times 10^{-6}$ & 10.72\% & 89.28\% &$-$                  & $5.35 \times 10^{-7}$& Yes \\
    H    & $5.21 \times 10^{-7}$ & $ 2.71\times 10^{-6}$ & 16.12\% & 83.88\% &$8.3 \times 10^{-8}$ & $1.04 \times 10^{-6}$& Yes \\
    \hline
    \end{tabular}
   \begin{list}{}{} 
   \item
    Mass (ejecta plus disk) gravitationally bound to the white dwarf, $\Delta M_{\rm WD}$, and total mass leaving the 
     binary system, $\Delta M_{\rm esc}$, 
     together with their fractions (in \%) over the 
     total ejecta plus disk masses, after collision of the disk with the nova ejecta. $\Delta M_{\rm SS, lost}$ is the mass lost from the system 
     by the secondary star in the interaction with the nova ejecta, while $\Delta M_{\rm SS, WD}$ is the mass of the secondary that is ultimately 
     accreted by the white dwarf.
\end{list}
    \end{table*}

\begin{figure*}[thb]
\centering
    \includegraphics[width=\textwidth]{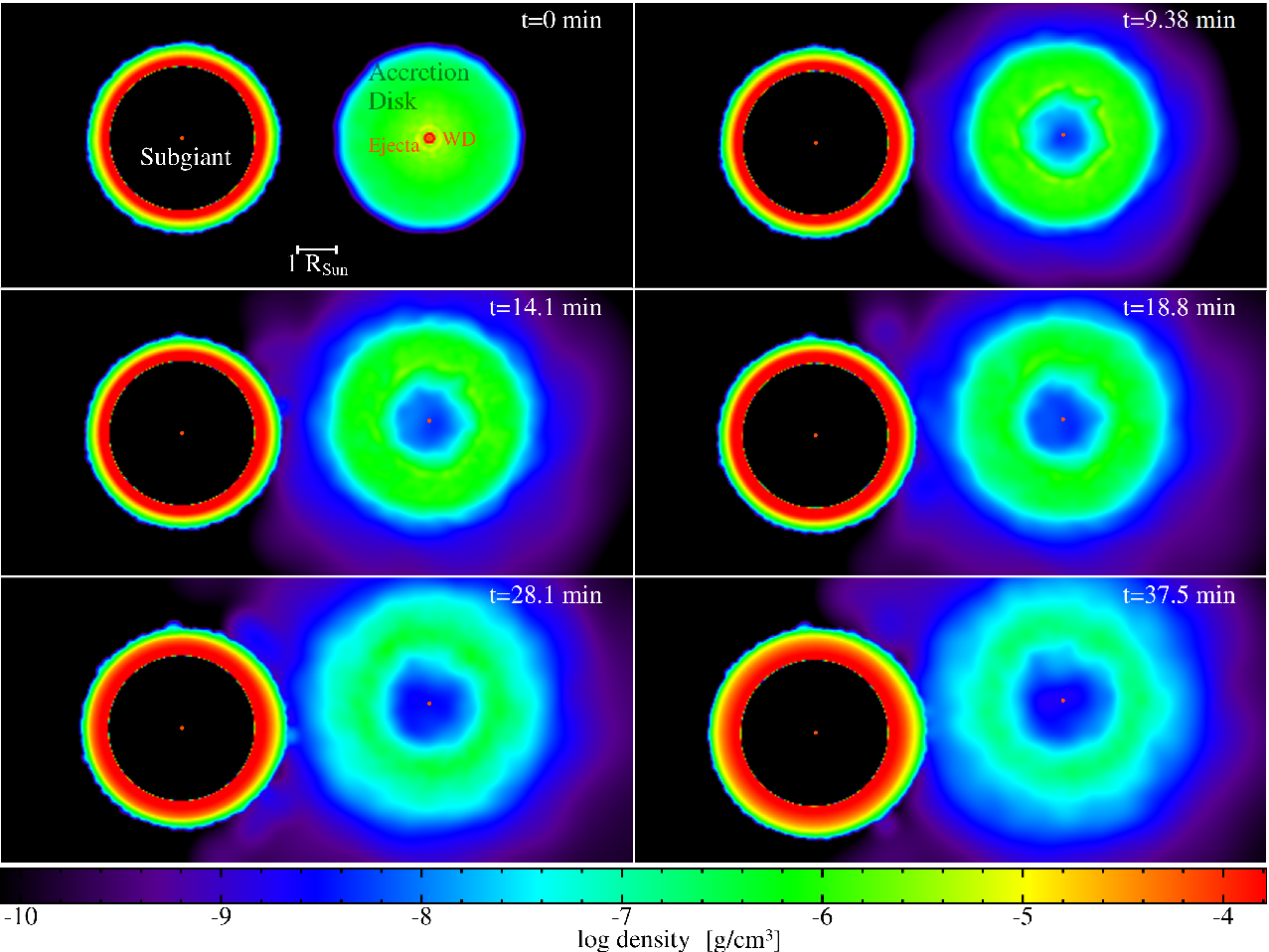}
\caption{Same as Fig. \ref{evolutionAA}, but for density plots corresponding to Model E. Note that in this model the accretion disk does not get totally swept up in the impact with the ejecta. 
A movie showcasing the time evolution of this model is available online.
Both the snapshots and the movie have been generated using the visualization software {\tt SPLASH} (Price 2007).}
     \label{evolutionEE}
      \end{figure*}

\subsection{Effect of the velocity of the ejecta}
We also explored the influence of the maximum velocity of the ejecta ($V^{\rm max}_{\rm ejecta}$) on the system dynamics, considering three different values: 
2000 km s$^{-1}$ (Model D), 5000 km s$^{-1}$ (Model C), and 10000 km s$^{-1}$ (Model A). The results, summarized in Table \ref{Series2b}, reveal a notable 
impact\footnote{Note that Models B and C, characterized by ejecta with 
similar momenta, result in similar amounts of mass loss from the binary system.}. An increase in the ejecta velocity increases the fraction of disk and ejecta mass that leaves the binary system: specifically, 91\% in Model D, 94\% in Model C, and 97\% in Model A. Simultaneously, the fraction of mass that remains gravitationally bound to the white dwarf is reduced, 
from 9\% in Model D to 6\% in Model C, and further to 3\% in Model A. 

The effect of the ejecta velocity on the fate of particles ejected by the subgiant companion, after the collision with the nova blast, is less clear: 
while the fraction of these particles ultimately accreted by the white dwarf tends to decrease with higher ejecta velocities, no clear trend is observed for the particles leaving the system 
(they achieve a maximum value of $\Delta M_{\rm SS, lost} = 2.75 \times 10^{-9}$ $M_{\odot}$ in Model C). It is noted, however, that some of values of $\Delta M_{\rm SS, lost}$ in Table \ref{Series2b} are very small and involve only a handful of particles, indicating a need for better resolution to shed light into these dependences.

\subsection{Effect of the density of the ejecta}
The impact of the specific density of the ejecta on the system dynamics has been assessed by comparing two models: Model A, characterized by an ejecta with a density profile (ranging from 
$\rho_{\rm ejec} = 7.9 \times 10^{-5}$ g cm$^{-3}$ at the outermost layers and $\rho_{\rm ejec} = 1.9 \times 10^{-4}$ g cm$^{-3}$ at innermost layers), and Model G, featuring a flared disk and constant ejecta density ($\rho_{\rm ejec} = 1.8 \times 10^{-4}$ g cm$^{-3}$). 
A closer examination of the results (Table \ref{Series2b}) reveals that the larger, constant density adopted in Model G leads to the deceleration of a greater number of particles during the collision with the disk. This results in a higher mass bound to the white dwarf star, rising from 3\% of the total disk and ejecta mass in Model A to 11\% in Model G. Conversely, the overall mass ejected from the binary system decreases, from 97\% in Model A to 89\% in Model G. 
Given that the disk mass is identical in the two models, the reported differences in results are ascribed to variations in the collision dynamics between ejecta and disk, 
primarily influenced by differences in the disk's density profiles. This underscores the importance of adopting realistic density profiles in such simulations.

The subsequent impact with the subgiant shows a decrease in the number of particles ejected from the secondary star that leave the system, dropping to zero in Model G. A small decrease is also observed in the number of particles from the secondary that ultimately get accreted by the white dwarf. Similar patterns are reported when comparing results from the simulations of V-shaped disks in Models H (disk with a density profile) and F (disk with constant density).

\subsection{Effect of the mass of the accretion disk}
The impact of the disk mass on the long-term evolution of the system has been analyzed by contrasting Models A and E, which feature $M_{\rm disk} = 2.1 \times 10^{-6}$ $M_\odot$ and $4.4 \times 10^{-6}$ $M_\odot$, respectively. Simulations reveal a significant influence on the system's fate. In Model A, characterized by a lighter disk, the nova blast fully disrupts the disk. Conversely, in Model E, the heavier disk partially survives the collision with the nova ejecta (see Fig. \ref{evolutionEE}). 
Moreover, a larger disk mass reduces the percentage of mass lost from the binary system (e.g., 70\% of the total disk and ejecta mass in Model E compared to 97\% in Model A), 
while increasing the amount that remains gravitationally bound to the white dwarf (30\% in Model E compared to 3\% in Model A; in both cases, mostly corresponding to disk material [$>$ 90\%]).
Furthermore, the larger amount of disk and ejecta that remains in the system in Model E, a fraction of which ultimately impacts the secondary star, results in a larger ejected mass from the subgiant. 

\subsection{Effect of the shape (geometry) of the disk}\label{GeoDisk}
The specific shape adopted for the accretion disk (flared vs. V-shaped) affects the cross-sectional area of the disk that is hit by the incoming nova ejecta. To quantify its effect, two models have been compared: Model A (flared disk) and Model H (V-shaped disk). The more extended V-shaped disk interacts with (and slows down) a larger fraction of the ejecta, reducing the overall mass lost from the binary system (84\% of the total disk and ejecta mass in Model H, compared to 97\% in Model A). Simultaneously, the mass remaining gravitationally bound to the white dwarf increases (3\% in Model A, 16\% in Model H).

The larger amount of disk and ejecta mass remaining in the system in Model H enhances the number of particle collisions with the secondary star. Consequently, this results in a higher mass ejected from the outer layers of the subgiant companion, increasing both the mass leaving the system ($\Delta M_{\rm SS, lost} = 8.3 \times 10^{-8}$ $M_{\odot}$ for Model H, $5.5 \times 10^{-10}$ $M_{\odot}$ for Model A) and the mass remaining bound to the white dwarf ($\Delta M_{\rm SS, WD} = 10^{-6}$ $M_{\odot}$ for Model H, $5.9 \times 10^{-7}$ $M_{\odot}$ for Model A).

Similar patterns are observed in the comparison between Model G (flared disk) and Model F (V-shaped disk), where the disks have a constant initial density. The effect of the different disk shapes is magnified, in the case of these models, by the survival of the disk
after the impact with the nova ejecta in Model F.

\section{Discussion}

It is notoriously difficult to see the disk in nova systems (S. Shore, private comm.). Several claims supporting disk disruption in nova outbursts (both classical and recurrent) have been made. According to Schaefer (2011), both the mass-accretion disk and the accretion stream appear to be blown away by the outburst in U Sco\footnote{See Drake \& Orlando (2010), 
for multidimensional simulations of the impact of the outburst on the accretion disk in U Sco.}. The accretion stream emerges 
on a freefall timescale\footnote{Observations suggest the reestablishment of an accretion stream 
about $8 - 15$ days post-explosion during the 2010 U Sco outburst (Schaefer et al. 2011, Mason et al. 2012). However, other interpretations have been proposed, in which the white dwarf photosphere expands largely at maximum and then begins to shrink down to $\sim 0.1$ R$_\odot$ when winds stop, allowing the disk to reappear from the white dwarf photosphere, some days after maximum.}. 
as the first discernible structure to reappear after the outburst, being limited by the wind that continues blowing away the infalling material. 
The outer edge of the reestablished accretion disk can only develop once  the stream approaches the white dwarf surface, coinciding with the dissipation of the wind. The development of this outer edge occurs on a timescale of the order of the orbital period of the system (Schaefer et al. 2011). Anupama et al. (2013) concluded that the flux variations of the N III feature and the intrinsic polarization observed in U Sco are most likely associated with the reforming accretion disk and/or stream. Similar conclusions were drawn by Ness et al. (2012), who interpreted the dipping in U Sco’s light curve as the result of occultations of the central source by high-density absorbing gas that is aligned along the trajectory of a reforming accretion stream. 
It is worth noting that the ejecta in U Sco may have a wide opening angle, potentially bypassing the disk. In fact, 
survival of the disk has been recently claimed in U Sco during the 2022 outburst (Muraoka et al. 2024),
suggesting the presence of an extended disk that reaches or even exceeds the Roche lobe size.  
Additional simulations incorporating such extended disks could help verify whether the critical mass ratios reported in this work also apply 
to these configurations.

Claims supporting disk disruption have been made in connection with other nova systems: Schaefer (2011) reported disk disruption during the 2000 eruption of the eclipsing RN CI Aql. Tofflemire et al. (2013) stated that the accretion disk in the RN T Pyx was disrupted or disturbed during the outburst. Henze et al. (2018), in the framework of the frequent explosions observed in the fast RN M31N 2008-12a, suggested the possibility of disk disruption in some of the past outbursts (those in which a lower mass-transfer rate led to a less massive disk). No clear evidence for a surviving disk was reported by Mason et al. (2021) in V1369 Cen. Azzollini et al. (2023) attributed the enhanced strength of the Balmer lines in the last spectra of the RN RS Oph to a newly reformed accretion disk around the white dwarf. Finally, it has also been suggested that in polar and intermediate-polar cataclysmic variables, systems characterized by strong and moderately strong magnetic fields, the magnetic field of the white dwarf could be strong enough to disrupt the inner part of the accretion disk (see, e.g., Wood, Abbott, \& Shafter 1992). 

The studies discussed in this section highlight the widely held view, backed by observational evidence, that nova ejecta can sometimes disrupt the accretion disk. While alternative interpretations have been proposed, the ongoing debate underscores the critical role of theoretical studies in elucidating the conditions that determine whether the disk is disrupted or survives during its interactions with classical and RN ejecta, reinforcing the significance of our study.

\section{Conclusions}\label{UScoConclusions}

The different masses, densities, and velocities of the ejecta inherent to recurrent and classical novae, as well as the different shapes and densities of the disks adopted, have motivated a reanalysis of the interaction between the ejecta, the accretion disk, and the companion star for the case of U Sco, a well-studied Galactic RN. To this end, we performed ten new 3D SPH simulations with rotation specifically designed to assess the impact of various parameters, including ejecta mass, velocity, and density, as well as the mass and geometry of the accretion disk, on the dynamical properties of the binary system. The key findings of our work are summarized as follows:
\begin{itemize}

\item An only very minor chemical contamination of the secondary star is anticipated in the U Sco case given the limited impact of nova ejecta particles on the subgiant.

\item Rotation has a marginal influence on the binding of disk plus ejecta particles to the white dwarf and on the overall mass lost from the  system in a wide orbit binary system like U Sco. However, 
 rotation seems to decrease the mass lost by the subgiant that ultimately becomes gravitationally bound to the white dwarf.

\item We investigated the conditions leading to disk disruption in a RN system. Six out of eight simulations reveal complete disruption and sweeping of the accretion disk orbiting the white dwarf star for models with flared disks and $M_{\rm ejecta}$/$M_{\rm disk} \geq 1$. In contrast, more massive disks ($M_{\rm ejecta}$/$M_{\rm disk} = 0.48$) and V-shaped disks with a (constant) high initial density partially survive the impact with the nova ejecta.  

\item A minor ejection of mass from the subgiant's outer layers is observed during the late-stage collision with ejecta and disk material, with some particles being ejected from the binary system and some accreted by the white dwarf.

\item Higher ejecta masses result in more mass escaping the binary system and less mass remaining gravitationally bound to the white dwarf. This is in sharp contrast with the results reported in Paper I and stems from the adoption of higher velocities for the ejecta, a distinctive feature of a RN system like U Sco. These velocities are high enough to compel a portion of both the disk and the ejecta to exceed the system's escape velocity following the collision.

\item The amount of mass leaving the binary system depends sensitively on the ejecta velocity, with higher velocities leading to more mass loss (and less mass bound to the white dwarf).

\item Uniform, denser ejecta cause a larger number of particles to decelerate  during the collision with the disk, resulting in more mass bound to the white dwarf and less overall mass ejected from the binary system compared to models with a density profile decreasing with distance to the white dwarf.

\item An increase in the accretion disk mass raises the likelihood of disk survival after a collision with the ejecta. Consequently, in models with more massive disks, the binary system  loses less mass, with more mass remaining gravitationally bound to the white dwarf.

\item The adoption of an extended V-shaped disk increases the fraction of ejecta slowing after collision, diminishing the overall mass lost from the binary system and increasing the mass bound to the white dwarf, compared to compact, flared disk geometries. This emphasizes the need for detailed disk
models for realistic  simulations of the long-term evolution of  RN (and classical nova) systems.

\end{itemize}

 It is important to note that the reduced allocation of particles for the low-density ejecta limits the scope of some of the presented results, which should be interpreted 
qualitatively. This type of simulations has not been attempted before, and we believe that the qualitative findings reported here will be valuable in guiding further research. 
High-resolution simulations are currently underway to validate these results. 

A potential extension of our work could involve incorporating an expanded envelope into the simulations, 
considering that both the accretion disk and the secondary star are actually embedded in the nova envelope.  

\begin{acknowledgements} 
We thank Steven N. Shore and Michael M. Shara for valuable comments. This work has been partially
supported by the Spanish MINECO grant PID2020-117252GB-I00, by the E.U. FEDER funds, by the AGAUR/Generalitat de Catalunya grant SGR-386/2021,
and by the Swiss Platform for Advanced Scientific Computing (PASC). The authors acknowledge support from the Center for Scientific Computing: sciCORE 
(\url{http://scicore.unibas.ch}) at the University of Basel, where part of the simulations were performed.
\end{acknowledgements}

\end{document}